\documentclass[twocolumn,showpacs,preprintnumbers,amsmath,amssymb,floatfix]{revtex4}

\usepackage{graphicx}
\usepackage{dcolumn}
\usepackage{bm}        
\usepackage{amsmath}
\usepackage{amsfonts}
\usepackage{amssymb}
\usepackage{appendix}

\def \simless {\mathbin{\lower 3pt\hbox{$\rlap{\raise 5pt 
              \hbox{$\char'074$}}\mathchar"7218$}}} 
\def \simgreat {\mathbin{\lower 3pt\hbox{$\rlap{\raise 5pt 
              \hbox{$\char'076$}}\mathchar"7218$}}} 
\def \dst{\displaystyle} 
 \def \sds {\strut\dst}

\def \vareps {\varepsilon}

\begin{document}

\title{Measurement of the total cross section \\
of uranium-uranium collisions at $\sqrt{s_{NN}}=192.8$~GeV}

\author{W. Fischer}
\email{Wolfram.Fischer@bnl.gov}
\author{A.J. Baltz}
\author{M. Blaskiewicz}
\author{D. Gassner}
\author{K.A. Drees}
\author{Y. Luo}
\author{M. Minty}
\author{P. Thieberger}
\author{M. Wilinski}
\affiliation{Brookhaven National Laboratory, Upton, NY 11973, USA}

\author{I.A. Pshenichnov}
\affiliation{Institute for Nuclear Research, Russian Academy of Sciences, 
   Moscow, Russia}

\begin{abstract}
Heavy ion cross sections totaling several hundred barns have been calculated 
previously for the Relativistic Heavy Ion Collider (RHIC) and the Large Hadron 
Collider (LHC). These total cross sections are more than an order of magnitude 
larger than the geometric ion-ion cross sections primarily due to Bound-Free 
Pair Production (BFPP) and Electro-Magnetic Dissociation (EMD). Apart from a 
general interest in verifying the calculations experimentally, an accurate 
prediction of the losses created in the heavy ion collisions is of practical 
interest for the LHC, where some collision products are lost in cryogenically 
cooled magnets and have the potential to quench these magnets. In the 2012 
RHIC run uranium ions collided with each other at $\sqrt{s_{NN}} = 192.8$~GeV 
with nearly all beam losses due to collisions. This allows for the measurement 
of the total cross section, which agrees with the calculated cross section
within the experimental error.
\end{abstract}


\maketitle

\section{Introduction}
The collisions of relativistic heavy ions have total cross sections as large
as hundreds of barns, primarily due to Bound-Free Pair Production (BFPP) and
Electro-Magnetic Dissociation (EMD). These cross sections have been previously
calculated for the Relativistic Heavy Ion Collider (RHIC)~\cite{RHIC1,RHIC2} 
and the Large Hadron Collider (LHC)~\cite{LHC1,LHC2}.

Secondary beams created by BFPP can limit the LHC heavy ion luminosity since 
they have a different charge-to-mass ratio than the primary beam and can 
be lost in cryogenically cooled magnets. The heat generated due to this loss 
can quench these magnets, when exceeding a threshold. Localized beam losses 
of a secondary beam generated by BFPP were observed in RHIC with Cu+Cu 
collisions~\cite{Bruc}. The secondary beams generated in Au+Au and U+U 
collisions in RHIC are within the transverse momentum aperture but outside
the longitudinal acceptance of the radio frequency (rf) buckets. These
secondary beams are eventually lost, but not in the same turn.

In 2012 RHIC operated for the first time with uranium ions, which have a 
prolate shape. Such collisions are interesting for several reasons. 
Collisions of uranium ions along their long axis can potentially create a 
quark-gluon plasma even denser than that from collisions of the more spherical
gold ions. Central collisions 
of uranium ions with the long axes parallel create elliptic flow~\cite{Acke} 
of secondary particles, but without the magnetic field generated by the ions 
passing each other when the elliptic flow is generated through partial overlap 
of the ions.

In uranium-uranium operation~\cite{Luo} with a low-loss magnetic 
lattice~\cite{Luo1} and 3-dimensional stochastic cooling in 
store~\cite{Blas1,Blas2,Blas3}, the beam loss of a well-tuned machine can be 
almost entirely attributed to luminous processes. In this situation the total 
cross section can be obtained from the observed beam loss rates.

In the following we present the calculations for the total U+U cross section 
for the RHIC operating energy, the main features of the  U+U operation during 
2012, the methodology for determining the total cross section experimentally, 
and the experimental result.


\section{Calculated total cross section}
The calculated U+U cross sections at RHIC are shown in Tab.~\ref{tab:xsec} 
together with the RHIC Au+Au and LHC Pb+Pb cross sections. The largest 
contributions to these cross sections come not from the nuclear overlap of the 
colliding ions, but from two electromagnetic processes, bound-electron 
free-positron pair (BFPP) production and electromagnetic dissociation (EMD) of 
the nucleus. Bound-electron free-positron pair production changes the charge 
of the ion causing it to fall out of the beam.  Electromagnetic dissociation 
removes at least one neutron from the ion, changing its mass and likewise 
causing it to fall out of the beam. 

Since the total U+U cross section measured here is obtained from beam loss 
rates, processes that do not lead to beam loss are obviously not included in 
the calculated cross sections. At RHIC energies the infinite range elastic 
Coulomb scattering cross section does not cause any notable deviation in the 
beam trajectories. The U+U cross section for free electron positron pairs 
in this experiment is huge, 64~kb calculated with the perturbative formula
of Racah~\cite{Rac37} (a formula which agrees very well with recent 
numerical calculations and is much more accurate than the formula of Landau and 
Lifschitz~\cite{LL34}). For Au+Au at RHIC the corresponding perturbative 
cross section is 36~kb. However, these free pair cross sections
are almost completely dominated by soft pairs, where the ions remain
intact, and they do not contribute to the present measured total
cross section.

\subsection{Bound-free pair production (BFPP)}
The calculated BFPP cross section for the process
\begin{equation}
 {^{238}\text{U}^{92+}} + {^{238}\text{U}^{92+}} \rightarrow
 {^{238}\text{U}^{92+}} + {^{238}\text{U}^{91+}} + \text{e}^{+} 
\end{equation}
is listed in Tab.~\ref{tab:xsec}. They are derived from the paper of Meier, 
Halabuka, Hencken, Trautmann, and Baur~\cite{Meie1}. In this work exact Dirac
wave functions are used for final state bound electrons, and higher shell
states are included.  Exact Coulomb distorted Dirac wave functions are also
used for the final state continuum positrons.  Calculations are perturbative,
and semi-classical straight line trajectories are assumed
for the ions.  However, higher order effects should be small: exact
calculations have been done in the ultra-relativistic limit for RHIC
Au+Au, and they show a modest reduction on the order of 3\% from
perturbation theory in the BFPP cross section to the dominant
lowest electron bound 1s-state~\cite{AJB97}.

The cross sections calculated in~\cite{Meie1} are compared in that paper with
other calculations in the literature and found to be in good agreement.
Because of the thoroughness and the convenient presentation of
the results in a useful form, we consider the~\cite{Meie1} results the current
state of the art for BFPP calculations. 

\begin{table}[htb]
\begin{center}
\caption{Calculated total cross sections for Au+Au and U+U collisions in RHIC,
and Pb+Pb collisions in the LHC. The total cross section is given by the sum
of BFPP, single EMD, and nuclear cross sections. The mutual 
EMD cross section is given for reference only.}
\label{tab:xsec}
\begin{tabular}{lcccc}
\hline\hline
collider          &     & RHIC   & RHIC  & LHC   \\
species           &     & Au+Au & U+U   & Pb+Pb \\
$\sqrt{s_{NN}}$    & GeV & 200    & 192.8 & 5520  \\
\hline
BFPP              & b   & 117    & 329   &  272   \\
single EMD        & b   &  94.15 & 150.1 &  215   \\
\hspace{2mm}{\it mutual EMD} & b   & {\it 3.79} & {\it 7.59} & {\it 6.2} \\
nuclear           & b   &   7.31 &   8.2  &   7.9 \\ 
\hline
total             & b   & 218.46 & 487.3  &  494.9 \\
\hline\hline
\end{tabular}
\end{center}
\end{table}

\subsection{Electro-magnetic dissociation (EMD)}
Ultraperipheral collisions of uranium nuclei at RHIC are characterized by the 
range of the impact parameter $b>2R_U$, where $R_U$ is the radius of uranium 
nucleus. It is measured along the major axis of a prolate uranium nucleus 
spheroid. The thickness of the diffuse nuclear boundary is included in $R_U$ to
avoid any overlap of nuclear densities at $b>2R_U$. In any case, we are 
ultimately interested in the sum of electromagnetic dissociation and 
dissociation from hadronic collisions of the nuclei as discussed below in 
Sec.~\ref{sec:hc}. There is a smooth transition as a function of decreasing 
impact parameter from electromagnetic to hadronic dissociation, and the total 
dissociation will be relatively insensitive to variations of $R_U$. This 
insensitivity with variations in $R$ was shown in an early calculation of the
Au + Au RHIC case~\cite{ZDC0}.

Strong nuclear forces are not involved in electromagnetic dissociation, but 
nuclei can be disintegrated by the impact of Lorentz-contracted Coulomb fields
of the collision partners. This phenomenon is well known as electromagnetic 
dissociation (EMD) of nuclei, see e.g. Refs.~\cite{PR2008,Pshenichnov:2011} and 
references therein. EMD events are classified into single and mutual 
dissociation events since experiments at heavy-ion colliders make it possible 
to register dissociation of nuclei either from one beam (with their collision 
partners frequently left intact) or simultaneously from both beams. This is  
contrary to collisions with strong interaction, which typically lead to 
mutual fragmentation of the colliding nuclei.      

The single EMD cross section
is a part of the total cross section measured in this work, while the mutual 
EMD cross section is relevant to luminosity measurements performed via the 
detection of correlated neutron emission in the Zero Degree Calorimeters (ZDC),
see Appendix~\ref{sec:app2} for details.

EMD cross sections for beam nuclei can be reliably calculated using the 
Weizs\"acker-Williams method of equivalent photons with measured total 
photoabsorption cross sections for these nuclei used as input. As demonstrated 
recently~\cite{Oppedisano:2011rx,AbelevALICE2012}, the RELDIS 
model~\cite{Pshenichnov:2011,Pshe1} describes well the absolute 
cross sections of neutron emission in single and mutual dissociation events 
resulting from ultraperipheral collisions of lead nuclei at the LHC. RELDIS is 
based on the Weiz\"acker-Williams method of equivalent photons and simulates 
their absorption by nuclei by means of the Monte Carlo method. 

In most events an excited heavy nuclear residue is created, which then 
evaporates neutrons. As predicted by RELDIS, only 3\% of single EMD events of 
lead nuclei at the LHC are without neutron emission. Therefore, the neutron 
emission EMD cross section for heavy nuclei serves as a good approximation of 
the total EMD cross section, but they are still distinguishable.
For U+U collision in RHIC the most common event is the single EMD process
\begin{equation}
 {^{238}\text{U}^{92+}} + {^{238}\text{U}^{92+}} \rightarrow
 {^{238}\text{U}^{92+}} + {^{237}\text{U}^{92+}} + \text{n}.
\end{equation}
The total single and mutual EMD cross sections calculated with RELDIS for 
Au+Au and U+U collisions in RHIC, and Pb+Pb collisions at the LHC are 
listed in Table~\ref{tab:xsec}.
A $Z^2$ factor - the square of the charge of the colliding nuclei - 
increases the total EMD cross sections in U+U collisions compared
to Au+Au collisions. This factor controls the intensity of the equivalent 
photon flux. In particular, as this factor appears twice in calculations  of
the mutual EMD cross section, the cross section calculated for U+U collisions 
at RHIC is twice as large as for Au+Au collisions with the same 
$\sqrt{s_{NN}}$. The mutual EMD cross section calculated for U+U collisions 
approaches the total nuclear cross section for such heavy and highly charged 
ion species. According to the RELDIS model, in addition to neutron emission, 
approximately $60$\% of electromagnetic excitation events of uranium nuclei 
lead to their fission.




\subsection{Hadronic collisions of nuclei}\label{sec:hc}
It is common to calculate the total cross section of hadronic collisions of 
nuclei (with overlap of their nuclear densities) by means of the Glauber 
model~\cite{Miller:2007ri}, which usually gives the total geometrical collision
cross section. A similar approach is calculating the total reaction 
cross sections for collisions of Pb nuclei with various targets in the 
abrasion-ablation model~\cite{Scheidenberger:2004xq}, which we also employ in 
the present work for calculating total cross sections. 

In addition to the distributions of nuclear densities of colliding nuclei, the 
elementary nucleon-nucleon cross section is a key input in Glauber-type 
calculations. Even soft fragmentation (e.g. a knock out of one or two nucleons)
leads to beam loss in a collider. Therefore, we use the total nucleon-nucleon 
collision cross section in our calculations instead of the inelastic one. We 
set the total nucleon-nucleon cross 
section to 52 and 95~mb for collisions at RHIC and LHC, respectively, according
to the compilation from Ref.~\cite{Antchev:2011vs}. This gives us the cross 
section for Au+Au at $\sqrt{s_{NN}}=200$~GeV of 7.31~b in full agreement with 
the total geometrical cross section in the optical limit, 7.28~b, from 
Ref.~\cite{Miller:2007ri}. The calculated total nuclear cross sections are
also shown in Tab.~\ref{tab:xsec}.






\section{RHIC in U+U operation}
The Relativistic Heavy Ion Collider (RHIC)~\cite{RHIC,Fisc} consists of two 
independent rings of 3.8~km circumference, named Blue and Yellow. It has 6 
interaction points (IPs) and presently provides collisions for the two 
experiments STAR at IP6 and PHENIX at IP8. Since 2000 RHIC has collided 
U+U, Au+Au, Cu+Au, Cu+Cu, d+Au, and polarized protons at 15 different 
energies. Recent upgrades have increased the heavy ion luminosity by
an order of magnitude through bunched beam stochastic cooling during
stores~\cite{Blas1,Blas2,Blas3}.

In 2012 uranium ions were collided for the first time in RHIC, which was also 
the first time for any hadron collider~\cite{Luo}. This was 
possible because a new Electron Beam Ion Source (EBIS)~\cite{EBIS} was 
recently commissioned that could provide enough intensity for collider 
operation. The magnetic lattice was selected to provide a large dynamic
aperture for on- and off-momentum particles with a beam envelope function
$\beta^*$ at the IP slightly larger than in previous years~\cite{Luo1}. 
Tab.~\ref{tab:beam} shows the main beam parameters typical for the highest
luminosity stores at the end of the 2012 running period. A total of 60 stores 
were provided for the experiments, with an average store length of 6.4~h.

\begin{table}[tbh]
\begin{center}
\caption{Main beam parameters during U+U operation. Values given are typical
for the highest luminosity stores at the end of the 2012 running period.
The initial value is at the beginning of stores, the value at 
$\mathcal{L}_{max}$ when the luminosity reached its maximum during the store, 
typically 1~h after the store started.}
\label{tab:beam}
\begin{tabular}{lccc}
\hline\hline
parameter                  & unit        & \multicolumn{2}{c}{value} \\
                           & & initial & \hspace*{2mm}at $\mathcal{L}_{max}$ \\
\hline
beam energy $E$            & \hspace*{-3mm}GeV/nucleon & \multicolumn{2}{c}{96.4} \\
number of bunches $n$      & ...         & \multicolumn{2}{c}{111} \\
bunches colliding at IP6 $n_{c6}$ & ...  & \multicolumn{2}{c}{102} \\
bunches colliding at IP8 $n_{c8}$ & ...  & \multicolumn{2}{c}{111} \\
bunch intensity $N_b$      & $10^9$      & 0.3  & 0.27 \\
beam current $I_b$         & mA          & 38   & 34  \\
normalized rms emittance $\vareps_{xy}$ & $\mu m$  & 2.25 & 0.40 \\
luminosity $\mathcal{L}$/IP & \hspace*{-3mm}$10^{26}$~cm$^{-2}$s$^{-1}$ & 3 & 9 \\
absolute beam loss rate $\dot{N}$   & 1000/s & 350 & 900 \\ 
relative beam loss rate $\dot{N}/N$ & \%/h   & 4   & 10 \\ 
\hline\hline
\end{tabular}
\end{center}
\end{table}

\begin{figure}[tbh]
\begin{center}
\includegraphics[width=85mm]{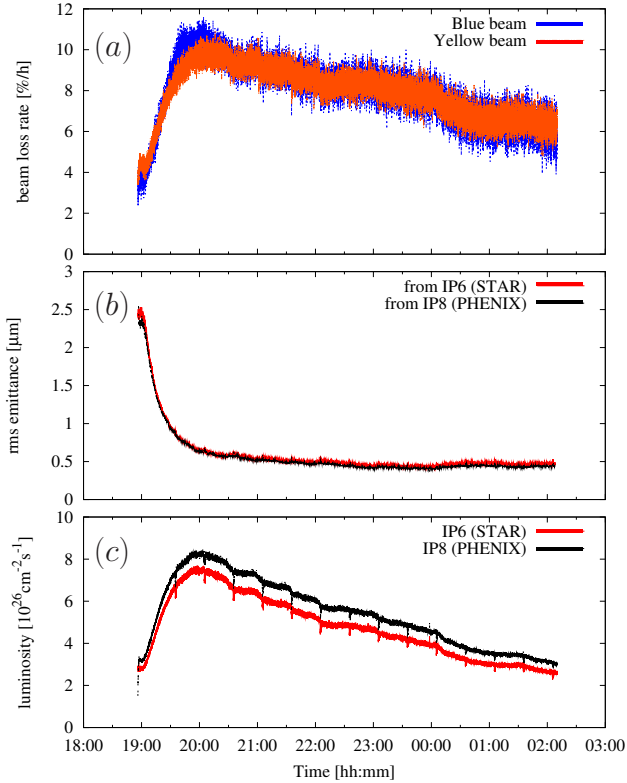}
\caption{(Color online) U+U store (fill number 16858) with relative beam loss 
rates (a), normalized rms emittances calculated from luminosity and intensity
(b), and luminosities at the STAR and PHENIX experiments (c). The 
reduction in the emittances and the corresponding increase in the luminosity 
are due to stochastic cooling during store.}
\label{fig:store}
\end{center}
\end{figure}

2012 was also the first year in which full 3-dimensional stochastic cooling
(i.e. horizontal, vertical, and longitudinal) was available in both rings. 
For uranium beams the cooling was so strong that the transverse emittances
were reduced by a factor of four (Fig.~\ref{fig:store}). The emittances are 
constant once the transverse intrabeam scattering growth~\cite{Piwi} rates 
and the cooling rates are in equilibrium. The peak luminosity increased by a 
factor of three (Fig.~\ref{fig:store}), and the average store luminosity 
by a factor of five. 

The 111 bunches leave 10\% of the circumference empty to allow for the abort 
kicker field strength to rise before arrival of the first bunch. With this 
abort gap all 111 bunches collided in the PHENIX experiment, but only 102 
bunches in the STAR experiment. This accounts for the visible difference in 
the STAR and PHENIX luminosities. 

The emittance shown in Fig.~\ref{fig:store} is averaged over all four 
transverse planes of both beams, and is calculated from the collision rate 
and the intensities. The low-loss lattice and the cooling resulted in beam 
losses (also visible in Fig.~\ref{fig:store}) nearly exclusively from burn-off 
through collisions. We discuss this in detail below.

\section{Total cross section measurement}
For round beams of the same size in both rings the time-dependent luminosity 
is given by~\cite{Furm1,Furm2}
\begin{equation}\label{eq:L}
\begin{split}
  \mathcal{L}(t) &= (\beta\gamma)\frac{n_c}{T_{rev}}\frac{N_{bB} N_{bY}}
	  {4\pi \vareps\beta^*}
          h(\sigma_s,\beta^*)  
\end{split}
\end{equation}
where $\gamma$ is the Lorentz factor, and $\gamma^2=1/(1-\beta^2)$,
$n_c$ is the number of colliding bunches ($n_{c6}=102$ for $\mathcal{L}_6$ at
STAR, $n_{c8}=111$ for $\mathcal{L}_8$ at PHENIX), 
$T_{rev}$ is the revolution time, $N_{bB},N_{bY}$ are the Blue 
and Yellow bunch intensities, $\vareps = \vareps_{x,B} = \vareps_{y,B}
= \vareps_{x,Y} = \vareps_{y,Y}$ is the normalized rms emittance, and 
$\beta^* = \beta^*_{x,B} = \beta^*_{y,B} = \beta^*_{x,Y} = \beta^*_{y,Y}$
the beam envelope function at the IP. The factor 
$h(\sigma_s,\beta^*)$, $\sigma_s$ being the longitudinal rms beam size, 
is not larger than and of order 1. It captures the hourglass and crossing 
angle effect. For Gaussian longitudinal distributions $h$ can be 
calculated~\cite{Furm1}. In the case of RHIC a numerical integration over the 
measured longitudinal particle distributions is necessary since the beams are
held in two radio frequency systems (harmonic numbers $h=360$ and 
$h=7\times 360$) and the bunches span several of the higher harmonic buckets 
at the end of a store.

In the case where all beam losses are given by 
the total U+U cross section $\sigma_{tot}$ we have 
\begin{equation}\label{eq:lossrate}
  \frac{dN_B(t)}{dt} = \frac{dN_Y(t)}{dt} = - \left[\mathcal{L}_6(t)+
  \mathcal{L}_8(t) \right] \sigma_{tot}
\end{equation}
where $N_B$ and $N_Y$ are the Blue and Yellow beam total intensities, and 
$\mathcal{L}_6(t)$ and $\mathcal{L}_8(t)$ the instantaneous luminosities at 
IP6 (STAR) and IP8 (PHENIX) respectively. The total cross section is then 
given by 
\begin{equation}\label{eq:sig}
  \sigma_{tot} 
    = - \frac{\dot{N}_{B,Y}(t)}
	{\left[\mathcal{L}_6(t)+ \mathcal{L}_8(t) \right]}
\end{equation}
and the systematic relative error by
\begin{equation}\label{eq:sigerr_rel}
  \frac{\Delta\sigma^{sys}}{\sigma_{tot}} = 
    \frac{\Delta\dot{N}_{B,Y}^{sys}}{\dot{N}_{B,Y}} + 
    \frac{\Delta\mathcal{L}_6^{sys} + \Delta\mathcal{L}_8^{sys}}
	 {\mathcal{L}_6 + \mathcal{L}_8}
\end{equation}
where $\Delta \dot{N}_{B,Y}^{sys}$ and $\Delta\mathcal{L}_{6,8}^{sys}$ are 
the systematic errors of the beam loss rate and luminosity, respectively. 
If timing errors can be neglected (see Sec.~\ref{sec:syserr}) we can use
\begin{equation}
  \frac{\Delta\dot{N}_{B,Y}^{sys}}{\dot{N}_{B,Y}} = 
  \frac{\Delta N_{B,Y}^{sys}}{N_{B,Y}}
\end{equation}
and the systematic error becomes
\begin{equation}\label{eq:sigerr}
 \Delta\sigma^{sys} = \sigma_{tot}\left( 
 \frac{\Delta N_{B,Y}^{sys}}{N_{B,Y}} + 
 \frac{\Delta\mathcal{L}_6^{sys}+\Delta\mathcal{L}_8^{sys}}
     {\mathcal{L}_6+\mathcal{L}_8}
 \right).
\end{equation}

We also note that if all beam losses are due to burn-off, both the transverse 
and longitudinal emittances are constant, and all bunches collide at IP6 and 
IP8 (i.e. $n_{c6}=n_{c8}=n_c$), the intensities and luminosities can be 
written as
\begin{equation}\label{eq:solL}
 N_{B,Y}(t) = \frac{N_{B,Y}(0)}{1+t/\tau} \;\;\; \mathrm{and} \;\;\;
 \mathcal{L}_{6,8}(t) = \frac{\mathcal{L}_{6,8}(0)}{(1+t/\tau)^2}
\end{equation}
where the time constant $\tau$ follows from Eq.~(\ref{eq:lossrate}) as
\begin{equation}\label{eq:tauL}
  \tau = \frac{N_{B,Y}(0)}
  {\left[\mathcal{L}_6(0)+\mathcal{L}_8(0)\right]
  \sigma_{tot}}.
\end{equation}

We need to obtain the beam loss rates $\dot{N}_{B,Y}$ and the luminosities 
$\mathcal{L}_{6,8}$ for the determination of the total cross section 
$\sigma_{tot}$ via Eq.~(\ref{eq:sig}). 
The beam loss rate is calculated from a time-dependent measurement of the 
total beam intensity with a Parametric Current Transformer (PCT, 
Appendix~\ref{sec:app1}). The total
intensity is reported every second and the loss rate is calculated as
the slope over an interval of 20~s. The luminosity is measured via the 
detection of neutron pairs in time-coincidence in the Zero Degree Calorimeter 
(ZDC)~\cite{ZDC2} (Appendix~\ref{sec:app2}). The luminosity is also reported 
every second.

\subsection{Luminous and non-luminous losses}
A principal limitation of the method used here comes from the fact that 
there can be beam losses from other processes than U+U interactions given by 
the total cross section $\sigma_{tot}$. In a situation with non-luminous 
losses, the use of equation (\ref{eq:sig}) will only give an upper limit for 
$\sigma_{tot}$ if these non-luminous losses are not subtracted. 

A number of tests are available to assess if there are non-luminous losses:
\begin{enumerate}
 \item The Blue and Yellow beam loss rates $\dot{N}_{B,Y}$ must be the same 
 if all losses are luminous.
 \item The Blue and Yellow beam loss rates $\dot{N}_{B,Y}$ must be proportional
 to the total luminosity $\mathcal{L}_6+\mathcal{L}_8$.
 \item The beam loss rates $\dot{N}_{B,Y}$ must be zero for non-colliding 
 beams.
 \newcounter{enumi_saved}
 \setcounter{enumi_saved}{\value{enumi}}
\end{enumerate}
These are necessary but not sufficient conditions for all beam losses to be 
luminous. We will use conditions 1 and 2 later to guide the data selection but 
we have no direct experimental data to test condition 3.

There are a number of processes that can lead to non-luminous beam losses:
\begin{enumerate}
 \item Intrabeam scattering 
\vspace*{-1mm}
 \item Residual gas elastic scattering
\vspace*{-1mm}
 \item Dynamic aperture and beam-beam effects
\vspace*{-1mm}
 \item Residual gas inelastic scattering
\end{enumerate}
In our discussion we need to separate emittance growth processes from beam 
loss processes. Emittance growth will eventually lead to beam loss once a 
limiting aperture is reached by the particles with the largest amplitudes.  
With stochastic cooling the emittance growth is reversed and an equilibrium 
emittance is reached for processes with growth times comparable to or smaller 
than the cooling time of about 1~h. Items 1-3 of the above list fall 
in this category. 

In some processes the particle amplitudes can be increased much faster than
the cooling time. Item 3 can and item 4 does fall in this category.
These processes are discussed in detail in Appendix~\ref{sec:app0} and 
summarized in Tab.~\ref{tab:calc}.
We must expect that there are a small number of particles lost 
through other processes than burn-off, and our analysis of the experimental
data needs to take this into account. The predominant process is inelastic
scattering on the residual gas, leading to beam loss rates of which are about
10\% of the total loss rate at the beginning of the store, and 3\% at the 
time of $\mathcal{L}_{max}$ (see Tab.~\ref{tab:calc} and Fig.~\ref{fig:fit}).

\begin{table}[tbh]
\begin{center}
\caption{Emittance growth and non-luminous beam loss for U+U collisions in 
RHIC with beam parameters given in Tab.~\ref{tab:beam}. Formulas are 
given in Appendix~\ref{sec:app0}. The values are shown for either the 
beginning of store or the time of $\mathcal{L}_{max}$, whichever gives the
larger emittance growth or beam loss. The emittance growth times are calculated
in the absence of cooling; cooling times are of order 1~h. Only the warm 
regions of the vacuum system are considered in the calculation of residual
gas effects. The cold regions with cryo pumping have very low particle 
densities.}
\label{tab:calc}
\footnotesize
\begin{tabular}{lcc}
\hline\hline
parameter & unit & value \\
\hline\hline
\multicolumn{3}{l}{\it warm vacuum sections} \\
length per ring $l_{w}$                         & m     & 652 \\
gas temperature                                 & K     & 300 \\
average pressure $\langle P\rangle$             & nTorr & 0.5 \\
static gas composition                          & ...   & \hspace*{-4mm}95\% H$_2$, 5\% CO \\
average $\beta$-functions $\langle\beta\rangle$ & m     & 115 \\ 
\hline
\multicolumn{3}{l}{\it emittance growth from intrabeam scattering} \\
transverse emittance growth time $\tau_{\vareps_n}$   & h & 0.44 \\
longitudinal emittance growth time $\tau_{\vareps_s}$ & h & 0.55 \\
\hline
\multicolumn{3}{l}{\it transverse emittance growth from residual gas elastic scattering} \\
emittance growth coefficient for N$_2$ & \hspace*{-5mm} s$^{-1}$Torr$^{-1}$ & 0.88  \\
N$_2$ equivalent pressure              & nTorr              & 0.5   \\
emittance growth time $\tau_{\vareps_n}$, at $\mathcal{L}_{max}$ & h & 49    \\
\hline
\multicolumn{3}{l}{\it beam loss from residual gas inelastic scattering} \\
coefficient for beam loss on N$_2$         & s$^{-1}$Torr$^{-1}$  & 800 \\
N$_2$ equivalent pressure                  & nTorr               & 0.1 \\
beam lifetime $\tau_N=N_{B,Y}/\dot{N}_{B,Y}$ & h                   & 498 \\
loss rate $\dot{N}_{B,Y}$, initial          & 1000~s$^{-1}$        & 19  \\
\hline\hline
\end{tabular}
\end{center}
\end{table}

\subsection{Measurement value and statistical error}
There were 60 physics U+U stores with an average length of 6.4 h 
(Tab.~\ref{tab:store}). As a first step we selected all stores which did not 
have any unusually high beam losses, and in which the Blue and Yellow beam 
loss rates $\dot{N}_B$ and  $\dot{N}_Y$ were approximately equal and 
proportional to the total luminosity $\mathcal{L}_6+\mathcal{L}_8$ for at 
least a period of the store. These were the last 50 of all 60 physics stores. 

\begin{table}[tbh]
\begin{center}
\caption{U+U store overview for $\sigma_{tot}$ determination.}
\label{tab:store}
\begin{tabular}{lcc}
\hline\hline
parameter                                           & unit & value \\
\hline
no of physics stores                                & ...  & 60  \\
average store length                                & h    & 6.4 \\
no of stores selected with $\dot{N}_B$ data         & ...  & 8   \\
no of stores selected with $\dot{N}_Y$ data         & ...  & 20   \\
total number of data points $(\dot{N},\mathcal{L})$ & ...  & 554k \\
\hline\hline
\end{tabular}
\end{center}
\end{table}

\begin{figure}[tbh]
\begin{center}
\includegraphics[width=60mm,angle=-90]{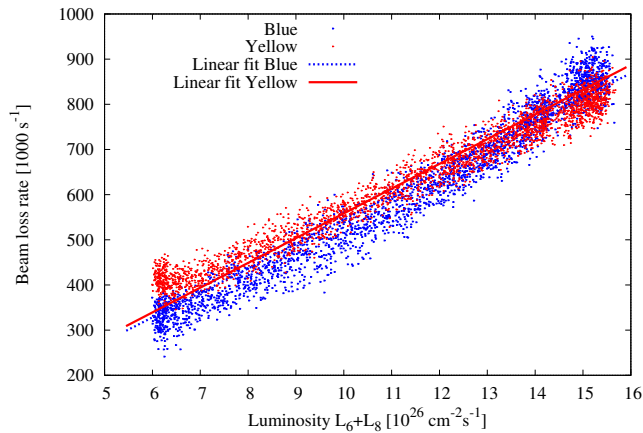}
\caption{(Color online) Blue and Yellow beam loss rates as a function of the 
total luminosity $\mathcal{L}_6+\mathcal{L}_8$ for fill number 16858. The 
units for the luminosity and beam loss rates are chosen so that the linear fit 
coefficient is returned in units of barn. Fitted values for this case are 
$\sigma_{tot} = (516.53\pm 0.21)$~barn for Blue, and 
$\sigma_{tot} = (549.61\pm 0.16)$~barn for Yellow. The error is the statistical
standard error.}
\label{fig:fit}
\end{center}
\end{figure}

For each of these stores a fitted value for $\sigma_{tot}$  was obtained by 
assuming that all losses are luminous, i.e. by fitting a straight line to
all pairs $(\dot{N},\mathcal{L})$ imposing the condition of a zero offset.
The fits for one store are shown in Fig.~\ref{fig:fit}, and the caption notes 
the standard statistical error for the fit. 

\begin{figure}[tbh]
\begin{center}
\includegraphics[width=59mm,angle=-90]{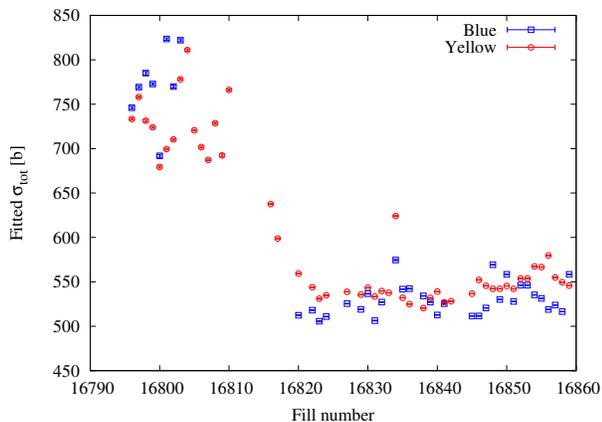}
\caption{(Color online) Fitted $\sigma_{tot}$ for all (partial) stores during 
which the beam loss rate was proportional to the total luminosity $\mathcal{L}_6
+\mathcal{L}_8$. All fills shown have longitudinal cooling. Vertical cooling
started in Yellow with fill 16816, and in Blue with fill 16820. Horizontal 
cooling started in Yellow with fill 16832, and in Blue with fill 16835.
In fill 16834 a cavity tripped leading to an increase in the momentum spread.
}
\label{fig:fills}
\end{center}
\end{figure}

Figure~\ref{fig:fills} shows the so fitted $\sigma_{tot}$ values for all 
pre-selected stores. The earlier stores (fill numbers 16796 to 16817) show a 
much larger fitted $\sigma_{tot}$ value than the later stores, indicating
that not all beam losses were luminous. We note that all pre-selected stores
have longitudinal stochastic cooling, and that transverse stochastic
cooling became available in the Yellow ring beginning with fill number 16816,
and in the Blue ring with fill number 16820, initially in the vertical plane
only. Beginning with fill number 16832 in the Yellow ring and 16835 in the
Blue ring the beams were cooled in all three dimensions. We further restrict
our data selections to fill number 16820 and larger. But even for those
fills we cannot be sure that all losses are luminous. Indications of this are 
the differences in the fitted $\sigma_{tot}$ from the Blue and Yellow beam
in the same store, and variations across stores. 

\begin{figure}[tbh]
\begin{center}
\includegraphics[width=60mm,angle=-90]{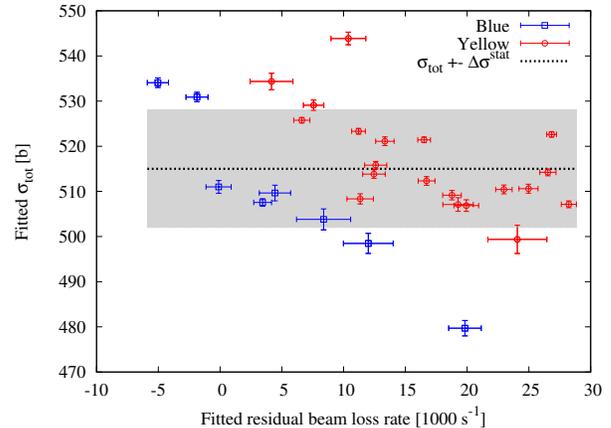}
\caption{(Color online) Fitted $\sigma_{tot}$ and fitted residual beam loss 
rates for selected stores. The errors of the individual data points are the 
standard errors from the linear fit. Shown is also the average of the 
$\sigma_{tot}$ values and the rms error of the combined individual data points 
as the gray area, giving $\sigma_{tot}=(515 \pm 13^{stat})$~barn.}
\label{fig:fills_wo}
\end{center}
\end{figure}

In a second step we fit $\sigma_{tot}$ to the $(\dot{N},\mathcal{L})$
data points for all stores with fill number 16820 and larger under the 
assumption that there are residual beam losses, i.e. with a nonzero
offset. In Fig.~\ref{fig:fills_wo} we show the fitted values for both 
$\sigma_{tot}$ and the residual beam loss rates, where we restricted the
data selection to data points with a residual beam loss rate not larger
than 30000~s$^{-1}$, about 10\% of the minimum and 3\% of the maximum beam 
loss rates observed (see Fig.~\ref{fig:fit}), to ensure that the 
$\sigma_{tot}$ determination is done with conditions of nearly all beam losses 
due to burn-off. 

Combining the individual data points in Fig.~\ref{fig:fills_wo}
yields the measurement value and statistical error as  $\sigma_{tot}=(515 
\pm 13^{stat})$~barn. We have calculated the statistical error simply as
the standard deviation of the $\sigma_{tot}$ distribution since this
distribution is much wider than the statistical errors for the individual
data points.

\subsection{Measurement systematic error}\label{sec:syserr}
If timing errors can be neglected, the systematic error is given by 
Eq.~(\ref{eq:sigerr}). We will now justify this assumption. The RHIC timing 
system is driven by a 5 MHz ultra-low noise temperature-controlled crystal 
oscillator 
with a relative frequency error of $\pm 10^{-9}$ to $\pm 10^{-10}$ per 
day~\cite{Wenz}. However, the timing error of reported 1~Hz signals, such as 
intensity and luminosity, have a much larger error. The combined systematic, 
periodic (due to clock beating), and random timing error can reach values of 
1~s, but typically does not exceed 0.5~s. Periodic and random errors with a 
symmetric distribution around a central value will only translate into 
statistical errors of the fitted total cross section (see Fig.~\ref{fig:fit}),
which we have taken into account already. 

Using the data shown in Figs.~\ref{fig:store} and \ref{fig:fit}, a systematic
timing error of 0.5~s is equivalent to a maximum systematic relative beam 
loss error of $(\Delta\dot{N}_{B,Y}/N_{B,Y})^{sys}_{max}=3.8\times 10^{-5}$ and 
a maximum systematic relative  luminosity error of 
$(\Delta\mathcal{L}_{6,8}/\mathcal{L}_{6,8})^{sys}_{max} = 3.3\times 10^{-5}$. 
These are much smaller than other systematic errors for the beam loss
rate and luminosity respectively (see below), and we therefore neglect timing 
errors.

The systematic error $\sigma^{sys}$ is then determined by the systematic 
errors of the beam intensity and luminosities. These are discussed in detail 
in Appendix~\ref{sec:app1} and Appendix~\ref{sec:app2} respectively.
The sources and contributions of all components are summarized in 
Tab.~\ref{tab:DI} and Tab.~\ref{tab:DL}, and, using Eq.~(\ref{eq:sigerr}) 
we obtain $\sigma^{sys}=22$~barn.

\begin{table}[tbh]
\begin{center}
\caption{Sources and contributions to the systematic errors in the beam 
intensity $\Delta N_{B,Y}^{sys}/N_{B,Y}$ (Appendix~\ref{sec:app1}).}
\label{tab:DI}
\begin{tabular}{lc}
\hline\hline
source                          & error \\
\hline
temperature variations          & 0.19\% \\
bunch pattern                   & 0.10\% \\
calibration error               & 0.15\% \\
PCT accuracy and readout drifts & 0.30\% \\
output noise                    & 0.01\% \\
\hline
total $\Delta N_{B,Y}^{sys}/N_{B,Y}$ (quadratic addition) & 0.40\% \\
\hline\hline
\end{tabular}
\end{center}
\end{table}

\begin{table}[tbh]
\begin{center}
\caption{Sources and contributions to the systematic errors in the 
luminosity $\Delta\mathcal{L}^{sys}/\mathcal{L}$ (Appendix~\ref{sec:app2} and
Ref.~\cite{Dree}).}
\label{tab:DL}
\begin{tabular}{lc}
\hline\hline
source                          & error \\
\hline
beam displacement               & 1.0\% \\
crossing angle                  & 2.0\% \\
intensity                       & 2.7\% \\
statistical                     & 1.7\% \\
\hline
total $\Delta\mathcal{L}^{sys}/\mathcal{L}$ (quadratic addition) & 3.9\% \\
\hline\hline
\end{tabular}
\end{center}
\end{table}

\section{Summary}
In U+U stores at $\sqrt{s_{NN}}=192.8$~GeV with 3D stochastic cooling nearly 
all beam losses are from burn-off and the total interaction cross section can 
be obtained from the observed beam loss rates as
\begin{equation}
 \sigma_{tot}^{meas} = (515 \pm 13^{stat} \pm 22^{sys} )
 \; \text{barn}
\end{equation}
with a combined statistical and systematic measurement error of 26~barn or 
5.0\%. The principal limitation of the measurement method are non-luminous
beam losses that are not accounted for. If these exist the measurement only
delivers an upper limit for $ \sigma_{tot}^{meas}$.
The calculated total cross section of $\sigma_{tot}^{calc} = 487.3\;
\text{barn}$ is smaller than the measured one by 28~barn or 5.4\%, a value 
close to the combined measurement error. 

\section{Acknowledgments}
The authors are thankful for discussions and support to J. Bergoz, S. Binello, 
R. Bruce, W. Christie, T. Hayes, X. He, M. Mapes, A. Marusic, K. Smith, and 
J. Jowett. Work supported by Brookhaven Science Associates, LLC under Contract
No. DE-AC02-98CH10886 with the U.S. Department of Energy.

\appendix
\section{Non-Luminous beam losses}\label{sec:app0}
\subsection{Intrabeam scattering}\label{sec:IBS}
Intrabeam scattering refers to small changes in the momenta of stored 
particles due to close encounters with other stored particles, leading to 
emittance growth~\cite{piw1}. In our case intrabeam scattering is counteracted 
by stochastic cooling, leading to a reduction of the emittance until an
equilibrium value is reached. 

Table~\ref{tab:calc} shows the emittance growth times
\begin{equation}
 \tau_{\vareps} = \left( \frac{1}{\vareps} \frac{d\vareps}{dt} \right)^{-1}
\end{equation}
in the absence of cooling for the transverse and longitudinal planes. 
Consistent with operational experience full transverse coupling is assumed and 
therefore the horizontal and vertical growth rates are equal. The calculation
is for the time when $\mathcal{L}_{max}$ is reached and the emittances are
close to their minimum value (Fig.~\ref{fig:store}).

We also estimate the lifetime of uranium beams due to intrabeam scattering 
in the presence of cooling. The emittance evolution is shown in 
Fig.~\ref{fig:store} where the asymptotic cooling time is about 1~h.
The collimator settings correspond to an rms emittance of $80~\mu$rad, and 
the simplest model is to assume the emittance distribution follows
\begin{equation} 
  \frac{\partial w(\vareps,t)}{\sds \partial t} = {\partial\over\partial 
  \vareps}\left( a\vareps w + M \vareps {\partial w 
  \over\sds\partial\vareps}  \right),
\label{mmb1}
\end{equation}
where $w(\vareps,t)$ is the time-dependent emittance distribution, $a$ is the 
cooling rate and $M$ is the diffusion rate. We have $a=1~{\rm h}^{-1}$ and 
$\vareps_a=M/a = 0.5~{\mu}$rad. If we put a collimator (an element that
ensures particle removal when a certain amplitude is exceeded) at a location
equivalent to an emittance of $\vareps_b = 80~{\mu}$rad the beam lifetime will 
be~\cite{piw1}
\begin{equation}
 \tau_{N} = 
 \frac{\vareps_a}{a \vareps_b}
 \exp\left(\frac{\vareps_b}{\vareps_a}\right) 
 \sim 10^{67} {\rm h}.
\end{equation}
Beam losses from intrabeam scattering with cooling can indeed be neglected.


\subsection{Residual gas elastic scattering}
We consider only the warm sections of RHIC since the residual gas density in 
the cold arcs is very low due to cryo pumping. The rms emittance growth 
time due to residual gas elastic scattering is~\cite{Mokh,Syph}
\begin{equation}\label{eq:1}
  \frac{1}{\tau_{\vareps_n}} = \frac{Z}{A}
  \frac{1}{\vareps_n}\frac{d \vareps_n}{dt}
  = \frac{(\beta\gamma)}{\vareps_n} \times
  \frac{1}{C}\int_0^C \beta(s)\theta_{rg}^2(s) ds
\end{equation}
where $Z$ and $A$ are the charge and mass number of the stored ion, 
$C$ is the circumference and $\theta_{rg}$ the rms scattering angle. For
protons, scattering on species with atomic number $Z_i$ and density $n_i$,
the scattering angle is~\cite{Mokh}
\begin{equation}\label{eq:2}
  \theta_{rg}^2(s) = \frac{4\pi r_p^2c}{\beta^3\gamma^2} 
  n_iZ_i(Z_i+1)\ln(183Z_i^{-1/3}),
\end{equation}
where $r_p$ is the classical proton radius and $c$ the speed of light. 
For molecular nitrogen N$_2$ at 300~K one has
\begin{equation}\label{eq:N2}
  \frac{1}{\tau_{\vareps_n}} \approx 0.88\, \mathrm{s}^{-1}\mathrm{Torr}^{-1}
  \frac{l_w}{C}\frac{\langle \beta P \rangle}{\vareps_n\gamma}.
\end{equation}
where $l_w$ is the length of the warm sections. For other residual gas species 
the N$_2$ equivalent pressure can be calculated according to Ref.~\cite{Mokh} as
\begin{equation}\label{eq:N2equiv}
\begin{split}
  P_{N_2 equiv}^{elast} &= \frac{2\times 10^{-3}}{P} \times \\
  &\phantom{=}
  \sum_i P_i
  \sum_j k_{ij} Z_{ij}(Z_{ij}+1)\ln(183 Z_{ij}^{-1/3}), 
\end{split}
\end{equation}
where $P_i$ is the partial pressure of gas molecules $i$, and $k_{ij}$ the 
number of species $j$ in the gas molecule $i$. Average $\beta$-functions 
of the warm sections and calculated emittance growth times are shown 
in Tab.~\ref{tab:calc}. 


\subsection{Residual gas inelastic scattering} 
Stored beam particles are lost after an inelastic collision with molecules of 
the residual gas in the beam pipe. The beam loss rate due to residual gas 
inelastic scattering of ions on molecular nitrogen N$_2$ at 300~K 
is~\cite{Fisc2,Mokh}
\begin{equation}\label{eq:blrg}
  - \frac{dN_{B,Y}}{dt} 
  \approx 800\,\mathrm{s}^{-1}\mathrm{Torr}^{-1} A^{2/3}
  N_{B,Y} \frac{\langle\beta\rangle}{C}\int_0^C P(s) ds,
\end{equation}
where $\langle\beta\rangle$ is the average $\beta$-function, $C$ the 
circumference, and $P(s)$ the $s$-dependent N$_2$ pressure. The nitrogen 
equivalent pressure can be calculated as~\cite{Mokh}
\begin{equation}
 P_{N_2 eqiv}^{nucl} = 0.0861\sum_i P_i \sum_j k_{ij}A_{ij}^{2/3}
\end{equation}
where $P_i$ is the partial pressure of gas molecules $i$, $k_{ij}$ are the 
number of species $j$ in the gas molecule $i$, and $A_{ij}$ the mass
number of species $j$ in molecule $i$. We consider only the warm beam pipe
regions, and neglect the cryogenically pumped cold beam pipe regions. The
relevant input data and calculated loss rate are shown in 
Tab.~\ref{tab:calc}. The calculated loss rate is for the beginning of the
store and has an error of approximately a factor of two, primarily due to the 
uncertainty in the pressure readings. The beam loss rate due to inelastic 
scattering decreases throughout the store with the decrease of the number of
stored particles.

\subsection{Dynamic aperture and beam-beam effects}
Orbit, tune, and chromaticity settings are chosen to minimize non-luminous
beam losses. However, actual machine conditions are not always reproducible,
and can drift with time, due to e.g. temperature changes. Beam losses are 
monitored continuously
during stores, and small parameter changes are periodically made to 
ensure that the machine operates at the minimum achievable loss rates. 
We disregard time-periods with beam loss rates higher than the consistently
low values established over a number of stores.

The motion of hadrons stored in a collider ring is well approximated by 
Hamiltonian mechanics (non-Hamiltonian effects were discussed above).
The particle motion in storage rings can be and typically 
is chaotic, which leads to emittance growth and possibly particle loss over 
the storage time~\cite{Seid,Seid1,Mess,Fisc3}. In addition to nonlinear elements
such as sextupoles, magnetic field errors in the main dipoles and quadrupoles, 
and the beam-beam interactions, there are parameter modulations that cause 
or enhance chaotic motion. 

Emittance growth effects due to nonlinear elements and parameter modulations
are generally smaller than the cooling time and will not lead to any relevant
losses (see Sec.~\ref{sec:IBS}). Chaotic particles can also be lost over
much smaller time scales than the cooling time~\cite{Fisc3}. But with cooling 
particles remain at small amplitudes, far away from the dynamic aperture,
and this is much less likely.

\section{Intensity measurement 
and systematic intensity error}\label{sec:app1}
The instruments to measure the number of ions present in each RHIC ring are 
two Bergoz Parametric Current Transformers (PCTs)~\cite{Berg}. These devices 
often referred to as DC Current Transformers (DCCTs) are 
sophisticated refinements of the basic fluxgate magnetometer~\cite{Evan} 
invented in the 1930s by Victor Vacquier at Gulf Research Laboratories,
and used in World War II to detect the proximity 
of submarines. The version now widely used for accelerator beam instrumentation
was invented in 1969 by K.B. Unser at CERN~\cite{Unse}  and further refined by 
J. Bergoz~\cite{Berg} and his collaborators. The PCTs measure all ions 
circulating, bunched and unbunched. The unbunched beam is never more than a
few percent of the total intensity, and since it is distributed over the full
circumference, its density is very low. It therefore does not contribute 
measurably to the luminosity.

It is important to note that this is a zero-crossing device in which the 
fluxes induced by the beam in very high permeability cores is compensated in 
a closed feedback loop by a current in windings that produce the opposite 
fluxes. In addition there are other windings that modulate the fluxes, causing
the cores to cross from saturation in one direction to saturation in the 
opposite direction. These transitions are detected by pickup coils and the lack
of symmetry (i.e. the presence of odd harmonics) of these signals is used as 
the error signal to close the feedback loop. The large dynamic range of these 
instruments (of order $10^7$) is mainly due to the fact that the flux in the 
cores is near zero no matter what the value is of the measured field. This 
also means that most measurement errors (except for noise) are only weakly
intensity dependent.

Rather than using the internal PCT instrument calibration, we use a precise 
external current source~\cite{Keith1} and digitize the output in a high 
precision Digital Volt Meter (DVM)~\cite{Keith2}. Both of these instruments 
are calibrated annually. 

The PCT calibration software in LabVIEW determines offset and slope values for 
the DVM readings as function of calibration current for both PCTs. These values
are then used to correct the readings and to arrive at values for the number 
of ions in each ring by taking into account their charge states and their 
revolution frequency. PCT calibrations are performed before each running period.

One issue investigated before starting the error evaluation was the 
placement of the coil used to inject the calibration current. The concern was 
that its different geometry compared to the beam may lead to a correction 
factor that may not have been considered. In fact, some facilities~\cite{Dena}
use straight calibration wires parallel to the beam. It was determined that this
is not an issue both experimentally and by consultation with 
experts~\cite{Blas,Berg1}. In the following we discuss one by one
the sources of error that contribute to the overall uncertainty in the beam 
intensity determinations.

\subsection{Temperature variations}
\begin{figure}[tbh]
\begin{center}
\includegraphics[width=80mm]{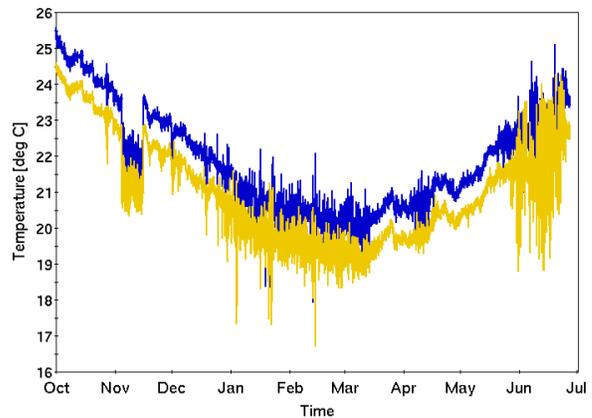}
\caption{(Color online) Blue and Yellow ring PCT core temperatures from 
October 2011 to June 2012, covering the period of the 2012 RHIC run.}
\label{figA1:2}
\end{center}
\end{figure}

According to the PCT User Manual the temperature coefficient for the 
electronics is $<0.1$~$\mu$A/K for the electronics but it is typically 
5~$\mu$A/K for the sensor head. There is no temperature regulation in the 
system nor are there corrections applied, even though the core temperatures 
are recorded. Figure~\ref{figA1:2} shows the temperature of the Blue and 
Yellow PCTs from October 2011 to July 2012. The maximum variation is 5K. 

Since the PCT calibration used is the one performed just before the 2012
running period when the temperature was highest, and the uranium operation took
place from 19 April to 15 May 2012, we use the full 5K excursion resulting 
in a current error due to temperature of 
5~$\mu$A/K~$\times$~5~K~$ =25$~$\mu$A. This translates into
an intensity error of 0.19\% (Tab.~\ref{tab:DI}).

\subsection{Bunch pattern influence}
\begin{figure}[tbh]
\begin{center}
\includegraphics[width=85mm]{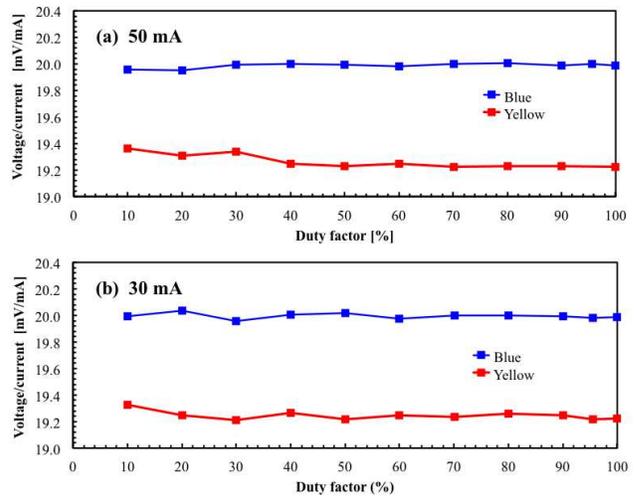}
\caption{(Color online) Measurement of the variation of the PCT output signal 
as a function of duty factor.}
\label{figA1:3}
\end{center}
\end{figure}

The variation of the PCT output signal as a function of the bunch pattern,
or duty factor, was measured by connecting a variable duty 
factor DC current supply to the calibration windings (Fig.~\ref{figA1:3}). 
This measurement shows that the fill pattern contributes less than 0.10\%
to the systematic intensity error (Tab.~\ref{tab:DI}).

\subsection{Calibration instrument errors}
Table~\ref{tabA1:1} shows examples for particular ranges for the accuracy in 
parts per million (ppm) of full scale for the DVM and the Current source that 
are used for the PCT calibrations.

We see that the only possibly relevant contribution to the error comes from 
the current source. It is 0.15\% of full scale if we add the drift over one 
year to the effect of a 5~K temperature uncertainty. This number
can be refined to take into account the use of other ranges and of currents 
that are not full-scale. Calibration errors yield an intensity error of
0.15\% (Tab.~\ref{tab:DI}).

\begin{table}[tbh]
\begin{center}
\caption{Accuracy and temperature coefficients of current calibration 
instruments~\cite{Keith1,Keith2}.}
\label{tabA1:1}
\begin{tabular}{lccccc}
\hline\hline
instrument            & range & 24 h  & 90 d  & 1 y   & temperature \\
                      &       &       &       &       & coefficient \\
                      &       & [ppm] & [ppm] & [ppm] & [ppm/K]     \\
\hline
Keithley 2000     &  10~V  & 15 & 20 & 30 & 2 \\
Keithley 220      & \hspace*{-2mm}100~mA & &  & \hspace*{-3mm}100,000 & 10,000\\
\hline\hline
\end{tabular}
\end{center}
\end{table}

\subsection{Zero and slope drift}
The PCT calibration is typically performed at the beginning of the running 
period and the question is by how much the calibration parameters can vary 
during the run. We have already considered the temperature effect above. Here 
we list other sources of errors that can affect the calibration taken from the 
specifications found in the PCT user manual:
\begin{itemize}
 \item Linearity error		$\pm 0.01$\%  $\pm$ zero error
 \item Zero drift (1 hour)	$< 1$~$\mu$A
 \item Zero drift (1 year)	$< 5$~$\mu$A at constant temperature
 \item Absolute accuracy	$> \pm 0.1$\%
\end{itemize}
We have compared some of these values with slope values for two instances 
where calibration parameters before and after a calibration were logged and 
one where calibrations were logged two days apart.

We see that all the slope differences between calibrations are largest for 
the MADC measurements which vary by an 
average of 0.14\% when the measurements are months apart and 0.036\% when 
they are 2 days apart. The DVM measurements are much better than could be 
expected from the current source specifications. Independent of the 
digitizer used the slope values always repeat within 0.2\%.

\subsection{Noise contributions}
\begin{figure}[tbh]
\begin{center}
\includegraphics[width=85mm]{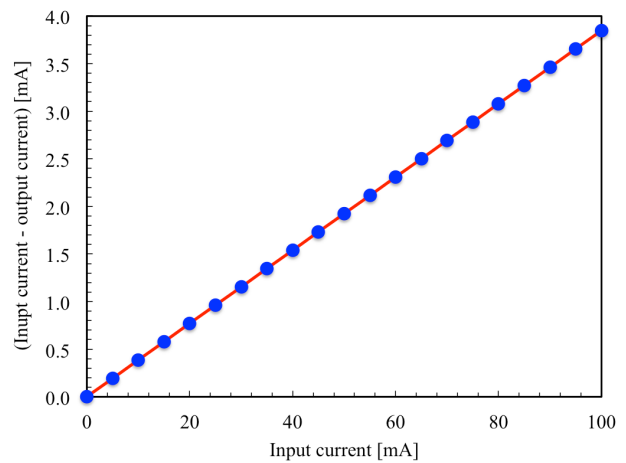}
\caption{(Color online) Difference between expected and observed DVM readings. 
The rms deviation from a straight line is 2.4~$\mu$A, not visible in this plot.}
\label{figA1:1}
\end{center}
\end{figure}

In the past, some discrepancies were observed when comparing expected DVM 
readings with actual readings obtained with a series of current source 
settings. These discrepancies were attributed to noise. Such measurements were 
repeated. Figure~\ref{figA1:1} shows the differences between expected and 
observed values and the quality of linear least square fit indicates that 
the discrepancies are mainly due to a calibration issue which is compensated 
for in the calibration procedure.

The error sources and values are summarized in Tab.~\ref{tab:DI}. 
For those errors that were obtained in units of current, such as the 
temperature correction, the values were converted to percentages by using an 
average beam current of 28~mA, typical for the uranium beams mid-store in 
the experiments (Tab.~\ref{tab:beam} shows the values at the beginning
of the store and at the time of the maximum luminosity).
Output noise contributes only 0.01\% to the systematic intensity error
(Tab.~\ref{tab:DI}).

\section{Luminosity measurement and 
systematic luminosity error}\label{sec:app2}

The luminosity is measured via the detection of neutron-pair coincidences in
the Zero Degree Calorimeter (ZDC)~\cite{ZDC0,ZDC2}. The effective cross
section for neutron-pair coincidence detection was measured to be
$\sigma_{nn} = 15.79$~barn for PHENIX, and $\sigma_{nn} = 15.84$~barn for STAR.
These values are in good agreement with the calculated cross sections 
given in Tab.~\ref{tab:xsec}. Indeed, by accounting for the fact that about 
97\% of the EMD events are accompanied by the emission of at least one neutron 
at each side, the value of 15.34~b is obtained as $(0.97 \times 0.97 \times 
7.59 + 8.2)$~b, providing that forward neutrons are always emitted in nuclear 
collisions.

A general outline of the error analysis of luminosity measurements is given
in Ref.~\cite{Whit}, and the detailed analysis of the U+U luminosity error in
Ref.~\cite{Dree}. The following effects were considered in the determination
of the luminosity error: (i) the need for a double Gaussian fit function for
the luminosity as a function of relative beam offset, (ii) the total intensity
measurement, (iii) the bunched intensity measurement, (iv) the fill pattern,
(v) the relative position measurement error, (vi) the crossing angle.
A summary of the error sources and contributions is provided
in Tab.~\ref{tab:DL}.

\end{document}